\documentclass[aps,prl,amsmath,amssymb,amsfonts,superscriptaddress,floatfix]{revtex4} 
\pdfoutput=1
\usepackage{amsmath,amsthm,amsfonts,amssymb}
\usepackage{graphicx}

\usepackage{color}
\usepackage{natbib,hyperref}
\usepackage{doi}

\DeclareMathAlphabet{\mathpzc}{OT1}{pzc}{m}{it}

\newcommand{\be}{\begin{equation}}
\newcommand{\ee}{\end{equation}}

\newcommand{\pplus}{\psi_{+}}

\begin{document}
\bibliographystyle{plainnat}

\title{The Short Path Algorithm Applied to a Toy Model}

\author{Matthew B.~Hastings}

\affiliation{Station Q, Microsoft Research, Santa Barbara, CA 93106-6105, USA}
\affiliation{Quantum Architectures and Computation Group, Microsoft Research, Redmond, WA 98052, USA}
\begin{abstract}
We numerically investigate the performance of the short path optimization algorithm on a toy problem, with the potential chosen to depend only on the total Hamming weight to allow simulation of larger systems.  We consider classes of potentials with multiple minima which cause the adiabatic algorithm to experience difficulties with small gaps.  The numerical investigation allows us to consider a broader range of parameters than was studied in previous rigorous work on the short path algorithm, and to show that the algorithm can continue to lead to speedups for more general objective functions than those considered before.  We find in many cases a polynomial speedup over Grover search.  We present a heuristic analytic treatment of choices of these parameters and of scaling of phase transitions in this model.
\end{abstract}
\maketitle

The short path algorithm is a recent quantum algorithm for combinatorial optimization\cite{Hastings_2018,weaker}. Consider a problem where one must optimize some function which depends on $N$ variables, each chosen from $\{-1,+1\}$; we write this function as an operator $H_Z$ which is diagonal in the computational basis for $N$ qubits.
The short path algorithm defines a  family of Hamiltonians depending on a parameter $s\in [0,1]$ by
\be
\label{Hsdef}
H_s=H_Z-sB (X/N)^K,
\ee
where $B,K$ are scalars and $X=\sum_i X_i$ with $X_i$ being the Pauli $X$ operator on qubit $i$.
Roughly, the algorithm is based on applying amplitude amplification to a subroutine that is defined as follows: one prepares the system in the state $\pplus$ defined to be ground state of $-X$, i.e., $\pplus$ is a state polarized in the $X$-direction.  One then evolves the Hamiltonian from $s=1$ to $s=0$.  Finally, one measures in the computational basis.
The evolution is done using a sequence of measurements (in some cases, only one measurement at the initial value of $s=1$ suffices) to allow it to be done to exponential accuracy.  In some cases, one can prove that the probability that the measurement in the computational basis gives the ground state is significantly larger than $2^{-N}$.  In such a case, amplitude amplification leads to a super-Grover speedup.

While the algorithm can in principle be applied to {\it any} combinatorial optimization problem, of course we expect that no speedup is possible for many potentials.  For example, if $H_Z$ is equal to $0$ on all basis states except for one computational basis state on which it is equal to $-1$, then no speedup over Grover is possible.
However, if the potential has some structure then some speedup is possible: previous results proved a speedup assuming that $H_Z$ was a weighted sum of products of Pauli $Z$ operators, each product being of the same degree $D$, and further assuming a bound on the low energy density of states.

However, it is also of interest to investigate this algorithm and other quantum algorithms numerically.  For example, the first paper on the adiabatic algorithm was a numerical study, and so a similar study is worthwhile for the short path algorithm.  Such numerical studies can be very useful because there are many algorithms whose performance is much better in practice than in a worst case theoretical analysis, or whose performance was only theoretically understood after they were already in widespread practical use (such as the simplex algorithm\cite{spielman2004smoothed}).  Since we do not yet have a working quantum computer capable of implementing many of these algorithms, simulation is the only tool to gain practical understanding.

In this paper, we give a first step to such a numerical investigation.  Indeed, the simulations are quite simple: first, one must investigate the gap on the parameter range $s\in [0,1]$ to verify that the evolution can be performed efficiently over this range.  We emphasize that in contrast to the adiabatic algorithm, we choose a {\it short path}.  The adiabatic algorithm tries to evolve from the Hamiltonian $H=-X$ to the Hamiltonian $H=H_Z$ adiabatically, which is expected to lead to super-exponential slowdown due to small gaps; see Ref.~\onlinecite{altshuler2010anderson} for general arguments and Ref.~\onlinecite{wecker2016training} for a small toy example.  We instead evolve only over a short range of $s$, choosing $B$ not to be too large so that the gap remains non-negligible.  Indeed, we will always keep this gap to be of order unity, i.e., to be lower bounded by some system-size independent quantity so that the scaling of this gap does not contribute to the runtime of the algorithm.  Second, one must compute the overlap of the ground state of $H_1$ with the state $\pplus$.  The probability that the subroutine succeeds is then proportional to the squared overlap, while after applying amplitude amplification we find that the algorithm runtime is proportional to the inverse of the absolute value of the overlap, multiplied by whatever time is required to the do the evolution from $s=1$ to $s=0$. Given that the gap in this interval is of order unity, then this time is only polynomial in $N$ for any ``reasonable" choice of $H_Z$, i.e., any choice with polynomially bounded entries.
The polynomial time to perform the evolution arises from the time required to perform the evolution over this interval to high accuracy use any of a number of different algorithms\cite{QSP,LC16,BerryEtAl2014,TS,BCK15}.

Let us define $B=bN$.  Then, the Hamiltonian is $H_Z-sb X^K/N^{K-1}$ so that for $K=1$ the value of $sb$ measures the strength of a
transverse field term.  At $s=1$, the Hamiltonian is $H_Z-bX^K/N^{K-1}$.
What we will find numerically in many cases is that while the gap tends to zero (either exponentially or super-exponentially in $N$) at certain values of $b$ or over a certain range of $b$, one can take $b$ to be slightly smaller (by some fixed, small $N$-independent amount $\delta$) than the critical value $b_{cr}$ at which it tends to zero.  In this case, the gap becomes $N$-independent and the overlap of the ground state with $\pplus$ is only weakly dependent on $\delta$.

So, in order to numerically study the short path algorithm, one must simply find the critical value of $b$, then choose some slightly smaller value and compute overlaps at this value. We describe this in more detail later.

In this paper, we consider a toy model in which $H_Z$ depends only on the total Hamming weight.  We choose this toy model so that we can simulate the system efficiently on a classical computer.  We do this by considering only a subset of states which are symmetric under permutation of qubits; the evolution under $H$ preserves this subspace.  Regarding each qubit as a spin-$1/2$ particle, these states are those states of total spin $N/2$.  There are $N+1$ such states, with each state being an equal amplitude superposition of computational basis states with given Hamming weight.  We consider an $H_Z$ with multiple minima so that there will be small gaps.

One interesting feature is that we are considering $H_Z$ to which the proofs of Ref.~\onlinecite{Hastings_2018,weaker} do {\it not} apply.  Those proofs considered an $H_Z$ which is a sum of monomials in Pauli $Z$ operators, with all monomials having the same degree.  The choices of $H_Z$ that we take here cannot be written in this fashion.  Thus, the numerical results here show that the short path algorithm can be useful outside the regime in which rigorous results are known.

We choose $H_Z$ with multiple well-separated minima.  As a result, the adiabatic algorithm becomes significantly slower than even brute force classical search.  We choose constants for which the speedup of the short path algorithm over a Grover search is only modest; this is done because it makes some of the numerical results easier to interpret.  Other choices of constants would lead to a much more significant speedup over Grover search.

\section{Toy Model}
As mentioned, the toy model consists of a potential $H_Z$ that depends only on the total Hamming weight, or, in physics language, on the total $Z$ polarization, i.e., on the value of $Z=\sum_i Z_i$.  The eigenvalues of $Z$ are $-N,-N+2,-N+4,\ldots,+N$.
We define the Hamming weight $w=(N-Z)/2$, so that a qubit with $Z_i=1$ corresponds to a bit $0$ while $Z_i=-1$ corresponds to a bit $1$.
As a toy model, we consider mostly the following simple piecewise linear form of $H_Z$ as a function of the Hamming weight $w$
\begin{eqnarray}
w\leq N/2+\delta_w & \quad \rightarrow \quad &H_Z=V_{max} \frac{w}{N/2+\delta_w}\\ \nonumber
w\geq N/2+\delta_w & \quad \rightarrow \quad & H_Z=\delta_V+(V_{max}-\delta_V)\frac{N-w}{N/2-\delta_w}.
\end{eqnarray}
That is, the function $H_Z$ increases linearly from $0$ at $w=0$ to $V_{max}$ at $w=N/2+\delta_w$, then it decreases linearly to $\delta_V$ at $w=N$.

Here we choose $V_{max}>0$ so that the function has two minima, one at $w=0$ and one at $w=N$.  This is done to produce more interesting behavior showing small gaps; if the function $H_Z$ were chosen instead to simply be proportional to $w$ then no small gaps appear for annealing.

\begin{figure}
\includegraphics[width=4in]{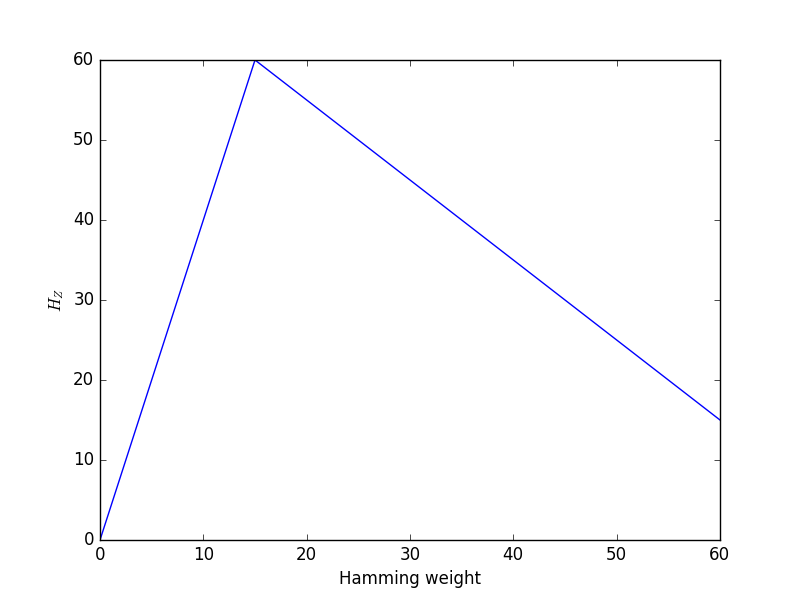}
\caption{$H_Z$ for $N=60,V_{max}=N,\delta_V=N/4,\delta_w=-N/4$.}
\label{figp}
\end{figure}

We pick $\delta_V>0$ so that the global minimum is at $w=0$.  However, we pick $\delta_w<0$ so that the minimum at $w=0$ has a smaller basin around it, i.e., so that the maximum of the potential is closer to $w=0$ than to $w=N$.
See Fig.~\ref{figp} for a plot of the potential $H_Z$.

We also later consider a slight modification of this potential, adding an additional small fluctuation added to the potential.  This is done to investigate the effect of small changes in the potential which, however, lead to additional minima and may lead to difficulties for classical annealing algorithms which may have trouble getting stuck in additional local minima which are created.  We do not perform an extensive investigation of this modification as our goal is not to consider in detail the effect on classical algorithms; rather, our goal is to understand the effect on the quantum algorithm.

The Hamiltonian of Eq.~(\ref{Hsdef}) depends upon a parameter $K$.  In previous work on the short path algorithm, this parameter was chosen to be an integer as that allowed an analytical treatment of the overlap using Brillouin-Wigner perturbation theory, and further it was chosen to be odd, again for technical reasons.  However, for numerical work, there is no reason to restrict ourselves to this specific choice of $K$ (on a gate model quantum computer, arbitrary values of $K$ may be implemented, using polynomial overhead to compute the exponent to exponential accuracy).
So, we investigate more general choices of $K$.
Whenever $K$ is taken to be an integer, we indeed will mean the Hamiltonian of Eq.~(\ref{Hsdef}).
However, if $K$ is taken non-integer, we instead consider the Hamiltonian
\be
\label{Hsdef2}
H_s=H_Z-sB (|X|/N)^K,
\ee
where the absolute value of $X$ means that we consider the operator with the same eigenvectors as $X$ but with the eigenvalues replaced by their absolute values.  We make this choice so that $|X|^K$ will still be a Hermitian operator; if instead we considered the operator $X^K$, then for non-integer $K$ this operator would have complex eigenvalues due to the negative eigenvalues of $X$.

Even with the choice of non-integer powers of $K$, the simulation of $H_s$ can still be performed efficiently as in Ref.~\onlinecite{Hastings_2018}
by implementing the operator $|X|^K$ in an eigenbasis of the operators $X_i$ where it is diagonal.

\section{Numerical and Analytical Results}
We now give numerical and analytical results.  Analytically, we consider some approximate expressions to locate the critical $b_{cr}$ at which the gap tends to zero as $N\rightarrow \infty$ and use this to help understand some of the scaling of the algorithms.

We consider three different algorithms and estimate their performance numerically.
These algorithms are the short path algorithm, the adiabatic algorithm, and a Grover speedup of a simple classical algorithm which picks a random initial state and then follows a greedy descent, repeating until it finds the global minimum.
In all cases, we will estimate the time as being $2^{CN}$ up to polynomial factors and we compute the constant $C$.  Smaller values of $C$ are better.  $C=1$ corresponds to brute force classical search while $C=0.5$ is a Grover speedup.

We consider three different choices of $H_Z$.  First, an ``extensive" $\delta_V$, i.e., one that is proportional to $N$.
This makes the situation much better for both the short path and adiabatic algorithm since the local minimum of the potential at $w=N$ is at a much larger energy than that of the minimum at $w=0$.
This situation is somewhat unrealistic, as in general we may expect a much smaller energy difference.  In this case, we are able to do the short path algorithm with $K=1$.
Second we consider $\delta_V=1$.  Here we find super-exponentially small gaps if $K=1$ and the location of the minimum gap tends to zero transverse field for $K=1$.  Since the location of the minimum gap tends to zero for $K=1$, we instead use larger $K$ for the short path algorithm so that the location of the minimum gap becomes roughly independent of $N$.

Up to this point, we find that the greedy descent is actually the fastest algorithm; perhaps this is no surprise since the potential is linear near each minimum so that so long as one is close enough, the descent works.
To get a more realistic situation, while still considering potentials that depends only on total Hamming weight,
our third choice of $H_Z$ has additional ``fluctuations" on top of the potential.  We take the potential $H_Z$ given before and add many small additional peaks to it so that the greedy descent will {\it not} work, instead getting trapped in local minima.  Thus, in this case, we do not consider at all the time for the greedy algorithm; however, we find that there is only a small effect on the performance of the short path algorithm.

The idea of adding fluctuations is similar to that of the idea of adding a ``spike"\cite{spike,Crosson_2016}, where one considers a potential that depends only on total Hamming weight which is linear with an added spike.  Here, however, we instead add many small peaks in the potential.  We do not in detail analyze the effect on the classical algorithm (for example, adding multi-bit-flip moves to the classical algorithm may enable to to avoid being trapped in minima).  Rather, the goal is just to consider the effect on the quantum algorithm.

We emphasize that, in contrast to previous work on spikes where the overall structure of the potential is linear with an added spike (so that without the added spike there is just one minimum), here we consider a potential which has multiple well-separated minima, even without any added spike.  The $\delta_w$ that we choose leads to only a very modest
speedup for the short path algorithm; different constants (in particular, making $|\delta_w|$ smaller) make the speedup more significant.  We choose the given value of constants as the interpretation of the numerical results was cleaner here.

\subsection{Results with Extensive $\delta_V$}
Here we consider extensive $\delta_V$.  In this subsection we use $K=1$ for the short path algorithm; later we will need larger $K$.

Consider the Hamiltonian $H_Z-bX$ for extensive $\delta_V$.
We can analytically estimate the location of the minimum gap and the value of the minimum gap as follows.
The minimum gap is due to an avoided level crossing.  To a good approximation, at small $b$, there is one eigenstate with its probability maximum at Hamming weight zero and another eigenstate with its probability maximum at Hamming weight $N$.  We can approximate these eigenstates by replacing the piecewise linear potential $H_Z$ by a linear potential that correctly describes the behavior near the probability maximum of the given eigenstate.

So, first we consider the Hamiltonian
$H=V_{max} \frac{w}{N/2+\delta_w}-bX$, which roughly describes the first eigenstate.
This Hamiltonian is equivalent to $\sum_i V_{max}(\frac{1-Z_i}{2})/(N/2+\delta_w)-bX_i$, which describes $N$ decoupled spins.
Each spin has ground state energy
\be
E_0\equiv \frac{V_{max}}{N+2\delta_w}-\sqrt{\Bigl(\frac{V_{max}}{N+2\delta_w}\Bigr)^2+b^2},
\ee
and so the total ground state energy is equal to $N$ times this.

The second eigenstate is roughly described by the Hamiltonian
$H=\delta_V+(V_{max}-\delta_V)\frac{N-w}{N/2-\delta_w}-bX$.  Again this describes $N$ decoupled spins, with ground state energy
per spin equal to
\be
E_1 \equiv \frac{\delta_V}{N}+
\frac{V_{max}-\delta_V}{N-2\delta_w}
-\sqrt{\Bigl(\frac{V_{max}-\delta_V}{N-2\delta_w}\Bigr)^2+b^2}.
\ee

We can estimate the value of $b$ where the gap is minimum by looking for a level crossing between $E_0$ and $E_1$.
This simple estimate is in fact highly accurate.
For example, for $V_{max}=N,\delta_V=N/4,\delta_w=-N/4$, using a Golden section search we find that $E_0=E_1$ at
$b=0.718070330\ldots$, while a numerical study of the exact solution with $N=40$ gave the crossing
at $b=0.718070335\ldots$, also using a Golden section search.

The important thing to note is that
the location of the level crossing occurs at a value of $b$ that is roughly independent of $N$ and that has a limit as $N\rightarrow \infty$ at some nonzero value of $b$.
At such a value of $b$, the level splitting is exponentially small in $N$.
A more careful treatment should also be able to estimate this level splitting quantitatively, i.e., one should be able to calculate the splitting scaling as $\exp(-c N)$ up to subleading corrections and to calculate the value of $c$.  We do not give this analytic treatment here and instead we are content to use numerical solution on finite sizes.

\begin{figure}
\includegraphics[width=4in]{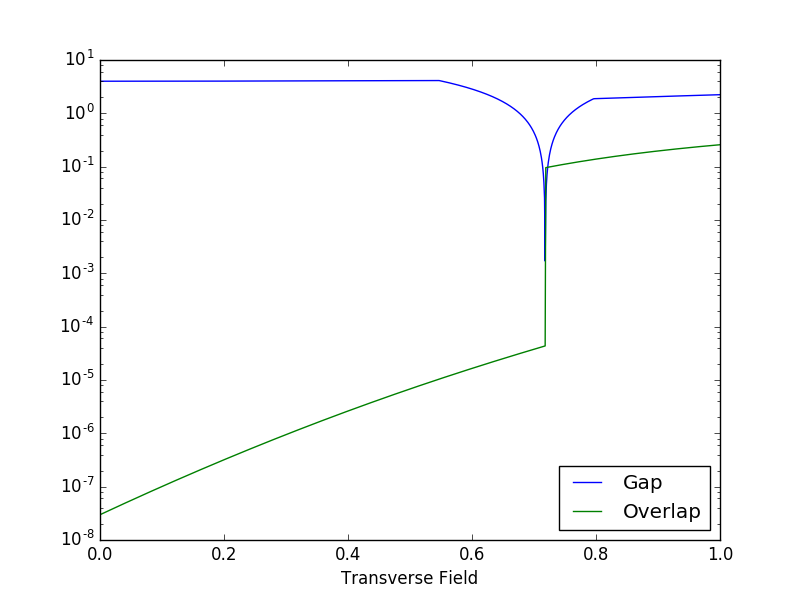}
\caption{Gap and overlap for $N=50,V_{max}=N,\delta_V=N/4,\delta_w=-N/4$.  Gap closing is not resolved finely enough to show the minimum gap on this figure.}
\label{fige}
\end{figure}

Fig.~\ref{fige} shows a plot of gap and overlap as a function of the transverse field strength $b$ for the case $N=50, V_{max}=N,\delta_V=N/4,\delta_w=-N/4$.  
As one can see, the overlap changes rapidly near where the gap becomes small.  The minimum gap was in fact $1.938\ldots\times 10^{-10}$, but the figure does not have enough resolution in the regime where the gap becomes small to see this small gap.  Let $b_{cr}$ be the value of the
transverse field where the gap tends to zero as $N\rightarrow\infty$.
One can also see that so long as we choose a value of $b=b_{cr}-\delta$ for some small $\delta>0$, then the exact choice of $\delta$ does not matter too much for the overlap.  For studying the short path algorithm, we picked $b=0.7$ for this specific set of parameters, which (for all sizes we studied) is comfortably far away from the gap closing.

We now estimate the time required for three different algorithms explained above.
To estimate the time for the short path, since we have picked a value of $b$ that is small enough that the gap is $N$-independent, the time is obtained from the scaling of the overlap.  We have computed the logarithm of the overlap to base $2$, and divided by $N$, for a range of sizes from $N=30$ to $N=50$.  For the adiabatic algorithm, we have computed the minimum gap, and taken the inverse square of the minimum gap as a time estimate, again using a range of sizes from $N=30$ to $N=50$.  
One finds that for both algorithms, the time estimate depends only weakly on the choice of $N$; it does get slightly worse as $N$ increases but the change is slow enough that we are confident that the numerical results provide a good estimate.

Note that in some cases, if one knows the location of the gap minimum to high accuracy and the gap grows rapidly away from the minimum, it is possible to use the techniques of Ref.~\onlinecite{roland2002quantum} to reduce the time so that it scales only as the inverse gap; in such cases the value of $C$ for the adiabatic algorithm is half that given here, however even if we halve the value of $C$ for the adiabatic algorithm the resulting $C$ is still larger than for other algorithms.

Of course, since the size of the Hilbert space is only {\it linear} in $N$ since we restrict to the subspace which is symmetric under permutation of qubits, it is possible to simulate systems of size much bigger than $N=50$.  However, since the minimum gap tends to zero exponentially, we run into issues with numerical precision when computing the gap at larger sizes, so we have chosen to limit to this range of sizes (if we were only interested in the short path algorithm, it would be possible to go to larger sizes since the overlap is not vanishing as rapidly).

To estimate the time for the classical algorithm, we compute the fraction of volume of the hypercube which is within distance $N/2-\delta_w$ of the all $0$ string.  This number gives the success probability; we take the inverse square-root of this number to get the time required using a Grover speedup.  This gives
$$2^{\frac{1}{2}\Bigl(
1-H(\frac{N/2-\delta_w}{N})
\Bigr)
},$$
where $H(p)=-p \log_2(p) - (1-p) \log_2(1-p)$ is the binary entropy function.

The results for the constant $C$ are $C=0.292\ldots$ for short path, $C=1.29\ldots$ for adiabatic and $C=0.094\ldots$ for the Groverized greedy algorithm.
Notably, the short path algorithm is significantly faster than the adiabatic algorithm.  However, the Groverized algorithm is the fastest.  In a later
subsection, we consider the effect of including ``fluctuations" to the potential which will prevent this simple Groverized algorithm from working but which have only a small effect on the performance of the short path algorithm.
Changing $\delta_w$ to $-3N_{spin}/8$ so that the basin near the global minimum becomes narrower, all algorithms slow down but the relative performance is similar:
$C=0.404\ldots$ for short path, $C=1.525$\ldots for adiabatic, and $C=0.22\ldots$ for the Groverized greedy.

\subsection{Results with $\delta_V=1$}
We now consider the case of $\delta_V=1$, keeping $\delta_w=-N/4$.
In this case, if we pick $K=1$, the location of the minimum gap tends to zero as $N\rightarrow\infty$.  To understand this, note that from
the decoupled spin approximation before, both eigenvalues decrease to second order in $b$.  The first eigenvalue is $0-c_1 N b^2+\ldots$ where the $\ldots$ denote higher terms in $b$ and where $c_1$ is some positive constant and the second eigenvalue is $1-c_2 N b^2+\ldots$, for some other constant $c_2$ with $c_2>c_1$.  These two eigenvalues cross at some value of $b$ which is proportional to $1/\sqrt{N}$.

The numerical results support this.  We considered a range of $N$ from $N=20$ to $N=60$.  Defining $b_{min}$ to denote the value of $b$ which gave the minimum gap, we found that $b_{min}*\sqrt{N}$ was equal to $1.41\ldots$ over this entire range.

As a result of this change in the location of the minimum gap, the value of the minimum gap is super-exponentially small.  This shifting in the location of the minimum gap is the mechanism that leads to small gaps as in Ref.~\onlinecite{altshuler2010anderson,Laumann_2015}.  The minimum gap is predicted to scale as $\exp(-c N \log(N))$ for some constant $c$.  We were not able to estimate this constant $c$ very accurately as the gap rapidly became much smaller than numerical precision.  However, even for very small $N$, the adiabatic algorithm becomes significantly worse than even a 
brute force classical search without Grover amplification; taking the time for the adiabatic algorithm as simply being the inverse square of the minimum gap, the crossover happens at $N \approx 10$.

This $N$-dependence of $b_{min}$ means that for the short path algorithm we cannot take $K=1$ and expect to get any nontrivial speedup.  However, we can take larger $K$.

\begin{figure}
\includegraphics[width=4in]{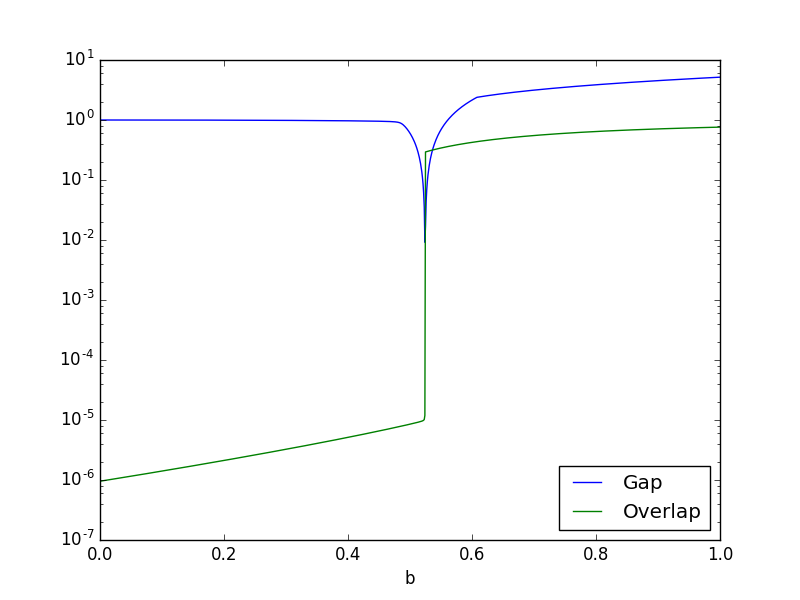}
\caption{Gap and overlap for $N=40$, $K=3$.}
\label{figK3}
\end{figure}

\begin{figure}
\includegraphics[width=4in]{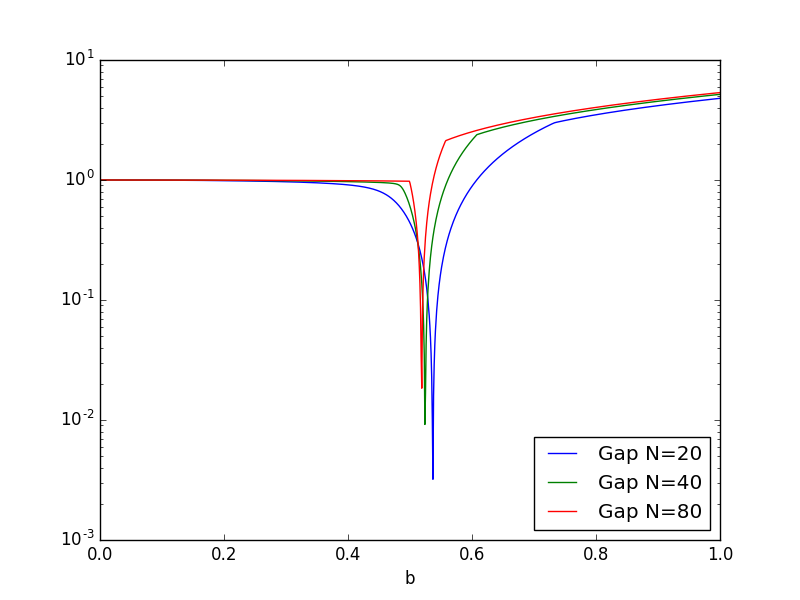}
\caption{Gaps for $N=20,40,80$, $K=3$.}
\label{figK3G}
\end{figure}

\begin{figure}
\includegraphics[width=4in]{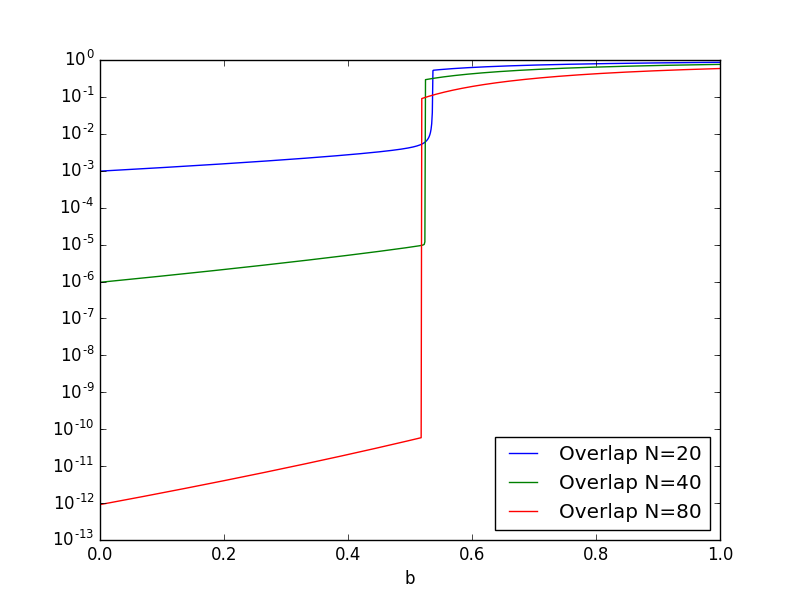}
\caption{Overlaps for $N=20,40,80$, $K=3$.}
\label{figK3O}
\end{figure}

Choosing $K=3$, the gap and overlap are shown
in Fig.~\ref{figK3} for $N=40$.  The figure does not have enough resolution to show the minimum gap accurately; the true minimum gap is roughly
$1.4\times 10^{-7}$.
However, now the location of the jump in overlap (and the local minimum in gap which occurs near that jump) become roughly $N$-independent.  Fig.~\ref{figK3G} and Fig.~\ref{figK3O} show the gap and overlap respectively for a sequence of sizes $N=20,40,80$.  The lines in Fig.~\ref{figK3G} all cross at roughly the same value of $b$.  The scaling behavior is less clear when the overlap is considered but is consistent with the jump in overlap becoming $N$-independent at large $N$ (the curve for $N=20$ shows some differences).

Using the short path scaling with $b=0.5$ (which is comfortably below the value at which the gap becomes small) we find a time $2^{CN}$ with $C=0.42\ldots$.  For smaller $\delta_w$, the value of $C$ reduces.

\begin{figure}
\includegraphics[width=4in]{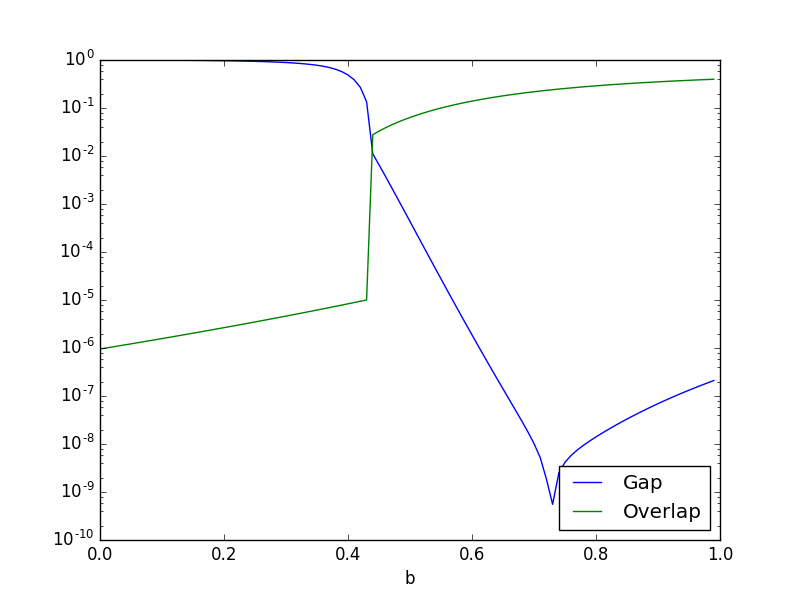}
\caption{Gap and overlap for $N=40$, $K=2$.}
\label{figK2}
\end{figure}

In contrast, for $K=2$, the location of the minimum gap does {\it not} have
an $N$-independent limit. 
First note that Fig.~\ref{figK2} shows a complicated behavior of the gap, which reduces rapidly near where the overlap jumps, but continues to stay small beyond that point.  The reason that the gap stays small even at large $b$ is that for $K=2$, the term $-X^K$ has a doubly degenerate ground state, one at $X=+N$ and one at $X=-N$.  Further, although the figure does not have the resolution to show it, the gap also becomes small near where the overlap jumps, i.e., there is another minimum of the gap for $b$ slightly larger than $0.4$.
So, there is a phase transition where the overlap jumps, with the gap becoming small there, and a degenerate ground state as $b\rightarrow \infty$.

\begin{figure}
\includegraphics[width=4in]{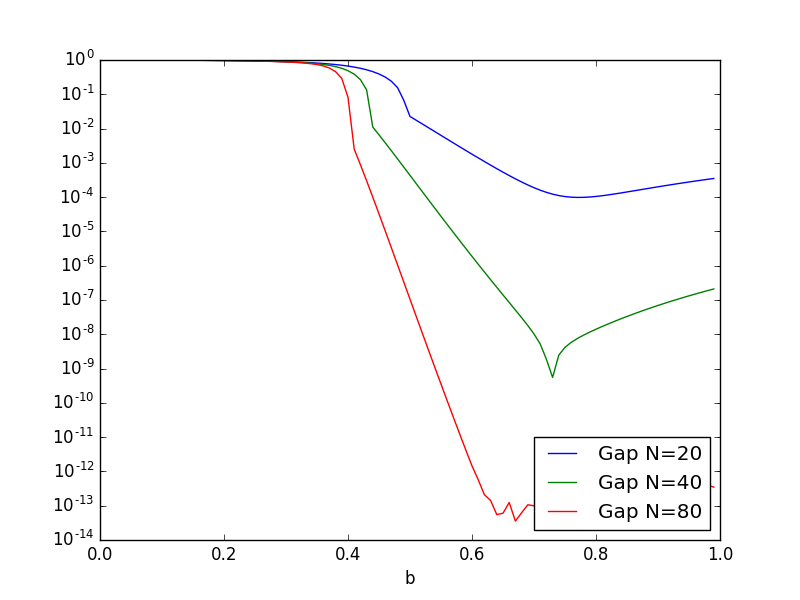}
\caption{Gaps for $N=20,40,80$, $K=2$.}
\label{figK2G}
\end{figure}

 Fig.~\ref{figK2G}
shows the gap for a sequence of sizes $N=20,40,80$.  Here we see that the curves shift leftwards as $N$ increases.
Previously, with the decoupled spin approximation we considered the Hamiltonian
$H=V_{max} \frac{w}{N/2+\delta_w}-bX
=\sum_i V_{max}(\frac{1-Z_i}{2})/(N/2+\delta_w)-bX_i$ to approximately describe the lowest eigenstate.
Now, we can try considering 
$H=V_{max} \frac{w}{N/2+\delta_w}-bX^2/N$.
This Hamiltonian does not describe decoupled spins.
Rather, it is equivalent to (up to an additive constant)
$$-\frac{V_{max}}{N+2\delta_w} Z-bX^2/N,$$
where $Z=\sum_i Z_i$.
We have $Z^2+X^2+Y^2=N(N+2)$ where $Y=\sum_i Y_i$.  So, for small $X,Y$ and large $N$ we can approximate $Z=N-(X^2+Y^2)/(2N)$.  Treating $X,Y,Z$ as classical variables (which becomes more accurate as $N$ becomes larger), we see that the minimum is obtained by taking $Y=0$ and then the Hamiltonian is only a function of $X^2$.  It is approximated by
$$X^2\cdot\Bigl(
\frac{V_{max}}{N+2\delta_w}\frac{1}{2N}-\frac{b}{N} 
\Bigr),$$
up to an additive constant.
This exhibits a phase transition as a function of $b$.  For $V_{max}=N$ and $\delta_w=-N/4$, this phase transition occurs at $b=1$.
For the other eigenstate, we consider the Hamiltonian
$H=\frac{V_{max}-\delta_V}{N-2\delta_w}Z-bX^2/N,$ up to an additive constant.
Using the same approximation $Z=N-(X^2+Y^2)/(2N)$, for $V_{max}=N$ and $\delta_w=-N/4$ this Hamiltonian
has a phase transition at $b=1/4$.
The plot shows values of the critical $b$ which are intermediate between $1$ and $1/4$, i.e., below the first phase transition but above the second.  Thus, it is possible that at large enough $N$, the leftward shift stops at $b=1/4$, so that the second eigenvalue reduces its energy due to the transverse field term but the first does not.  We leave this for future work.

One might try to consider $K$ intermediate between $2$ and $3$ to see if some further speedup is possible.  We leave this for the future.

\subsection{Added Fluctuations}
We now consider the effect of adding fluctuations.  We again took $\delta_V=1,\delta_w=-N/4$ as in the previous subsection.  However, we modified the potential by adding on an additional value $f$ if the Hamming weight was equal to $1$ mod $2$.  This choice was chosen so that for $K=3$ (the case we took) the transverse field term connects computational basis states which differ by $1$ mod $2$ so that the added fluctuations will have some effect.  We picked a large value of $f$, equal to $N/4$, so that classical simulated annealing would have exponentially small probability to move over the fluctuations.  We picked $N$ even so that the added fluctuations have no effect on the values of $H_Z$
 at $w=0$ and $w=N$.
 
\begin{figure}
\includegraphics[width=4in]{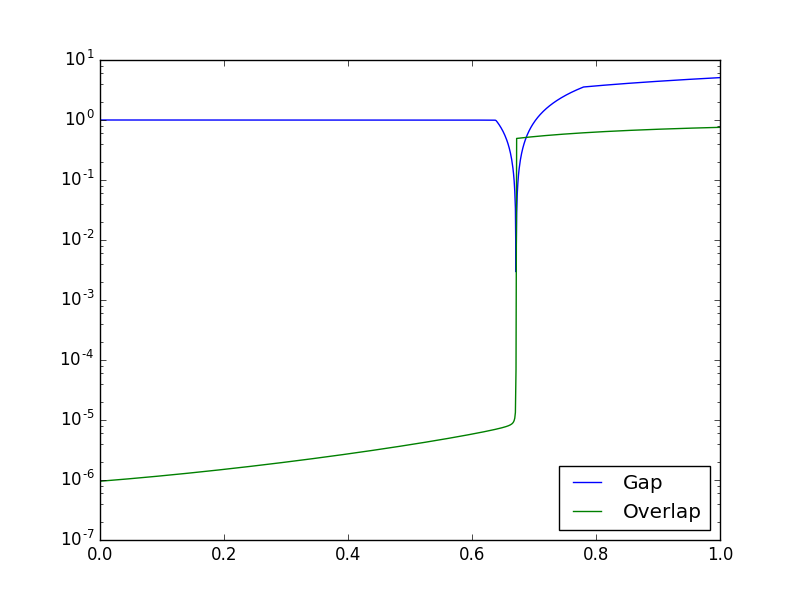}
\caption{Gap and overlap for $N=40$, $K=3$, with added fluctuations.}
\label{figf}
\end{figure}

\begin{figure}
\includegraphics[width=4in]{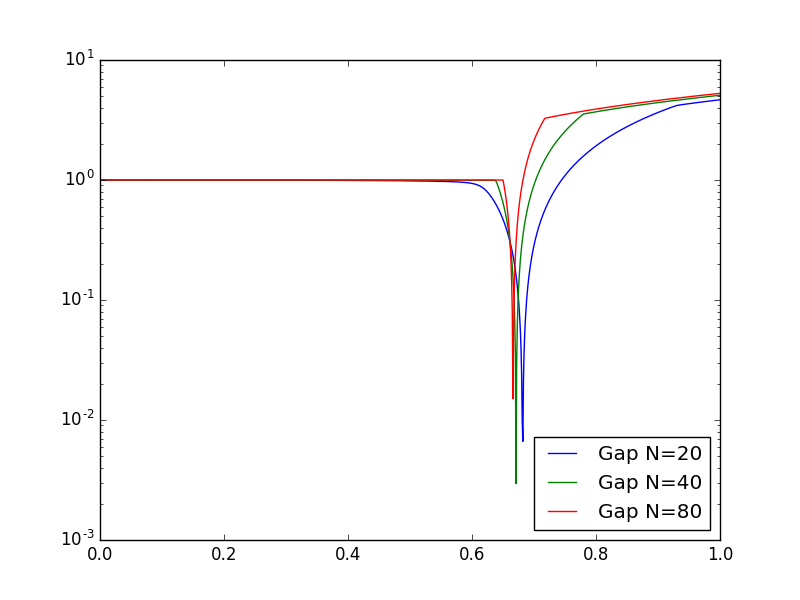}
\caption{Gaps for $N=20,40,80$, $K=3$, with added fluctuations.}
\label{figfg}
\end{figure}

As seen in Fig.~\ref{figf}, the shape of the gap and overlap is similar to the case without fluctuations, except that there is an overall rightward shift.  As seen in Fig.~\ref{figfg}, the location of the small gap is again roughly $N$-independent.
Because of the rightward shift we are able to take a larger value of $b$ in the short path than we could without fluctuations; however, the overlap at this value of $b$ is roughly the same as without fluctuations.  Thus, we find almost the same time scaling as before, in this case $C=0.43\ldots$

\section{Quantum Algorithms Projected Locally}
As expected, the adiabatic algorithm has problems with multiple minima.  Depending on the values of $\delta_V,\delta_w$, this can lead to either exponentially small gaps (sufficiently small that in many cases the algorithm is slower than brute force search) or even super-exponentially small gaps.  Thus, we suggest that it may be natural to consider the following modification of the adiabatic algorithm.
Indeed, this modification could be applied to the short path algorithm as well, though we explain it first for the adiabatic algorithm

We explain this modification for {\it arbitrary} functions $H_Z$ rather than just the specific choices here which depend only on $w$.
Define as a subroutine an ``adiabatic algorithm projected locally" as follows: pick some bit string $b$ and some distance $d$.
Then, consider the family of Hamiltonians
$sH_Z-(1-s)X,$
restricted to the set of computational basis states within Hamming distance $d$ of bit string $b$.
That is, defining $\Pi_{d,b}$ to project onto that set of computational basis states, we consider the family
$$\Pi_{d,b} \Bigl(sH_Z-(1-s)X\Bigr)\Pi_{d,b}$$
restricted to the range of $\Pi_{d,b}$.
One applies the adiabatic algorithm to this Hamiltonian for the given choice of $d,b$ and (hopefully) if $d$ is small enough, no small gap will appear so that the adiabatic algorithm will be able to efficiently search within distance $d$ of bit string $b$.

To implement the projector $\Pi_{d,b}$, one first must decide on a representation for the set of states within Hamming distance $d$ of $b$.
The simplest representation is an overcomplete one: simply use all bit strings of length $N$ (other representations are possible but they make the circuits more complicated).  Then, the projector $\Pi_{d,b}$ can be computed using a simple quantum circuit: given a bit string, exclusive-OR the bit string with $b$ and then use an adder to compute the Hamming weight of the result, then compare the result of the addition to $d$, and finally uncompute the addition and exclusive-OR.  The transverse field terms $X_i$ in the adiabatic algorithm can then be replaced with the corresponding terms $\Pi_{d,b} X_i \Pi_{d,b}$ (to do this in a gate model, once $\Pi_{d,b}$ is computed on a given basis state, one can use it to control the application of $X_i$ and then uncompute $\Pi_{d,b}$, so that now the Hamiltonian commutes with $\Pi_{d,b}$).

There are a couple ways to
prepare the initial state which is a ground state of $-\Pi_{d,b} X \Pi_{d,b}$.  We very briefly sketch this here, leaving the details for future work.  One is to note that the amplitudes in such a state depend only on the Hamming distance from $b$ (states closer to $b$ may have larger amplitude than those further, for example) and such a state is a matrix product state (applying an arbitrary ordering of qubits to regard them as lying on a one-dimensional line) with bond dimension at most $d$ and so can be prepared in polynomial time.  More simply, one can adiabatically evolve to such a state by considering a path of Hamiltonians $V(s)-X$.  Here, the potential $V(s)$ is diagonal in the computational basis and depends on a parameter $s$.  We let the potential $V(s)$ equal $0$ for $s=0$, and (as $s$ increases) the potential gradually increases at large Hamming distance from $b$, while keeping $V$ equal to $0$ on states within Hamming distance $d$ of $b$.  One may find a path of such potentials so that the amplitude at distance greater than $d$ from $b$ becomes negligible and such that the gap does not become small (note that here we choose $V$ to increase monotonically with increasing Hamming distance from $b$, rather than having multiple minima, to avoid any small gaps).  

Then, given this subroutine, we consider the problem of minimizing $H_Z$.
We consider this as a decision problem: does there exist a computational basis state such that
$H_Z$ has some given value $E_0$?
So, one can define an algorithm which takes a given $d$ and then chooses $b$ randomly, and applies the adiabatic algorithm projected locally for the given $d,b$ in an attempt to find such a computational basis state; if $E_0$ is the minimum value of $H_Z$ within distance $d$ of bit string $b$, and if no small gap arises, then the adiabatic algorithm will succeed.

Finally, one takes this algorithm with random choice of $d$ and applies amplitude amplification to it.  Thus, in the case that $d=N$, the choice of $b$ is irrelevant and we have the original adiabatic algorithm, while if $d=0$ the algorithm reduces to Grover search.

One can do a similar thing for the short path: choose a random $d$, restrict to the the set of computational basis states within Hamming distance $d$ of bit string $b$, and apply the short path algorithm on that set.  Finally, apply amplitude amplification to that algorithm rather than choosing $d$ randomly.

It may be worth investigating such algorithms.  Interestingly, it is possible to study this algorithm efficiently on a classical computer for choices of $H_Z$ which depend only on total Hamming weight, such as those considered above.
To do this, suppose that bit string $b$ has total Hamming weight $w_b$.  Then, without loss of generality, suppose that bit string $b$ is equal to $1$ on bits $1,\ldots,w_b$ and is equal to $0$ on bits $w_b+1,\ldots,N$.
Then, the Hamiltonians and projectors considered are invariant under permutation of bits $1,\ldots,w_b$ and under permutation of bits $w_b+1,\ldots,N$, so that we may work in the symmetric subspace under both permutation.
The basis vectors in this symmetric subspace can be labelled by two integers, $w_1,w_2$, where $w_1=0,\ldots,w_b$ is the total Hamming weight of bits $1,\ldots,w_b$ and $w_2=0,\ldots,N-w_b$ is the total Hamming weight of bits $w_b+1,\ldots,N$.
Then, this gives us a basis of size $O(N^2)$ and hence we can perform the classical simulation efficiently.

However, we suspect that while this may be useful for the very simple piecewise linear potentials considered here, it will probably not be useful for more general $H_Z$.  If there are many local minima (so that for any $b$ which is proportional to $N$ there are many comparable local minima in that basin), then there will probably still be a slowdown for either algorithm.

\section{Discussion}
We have considered the short path algorithm in some toy settings.  We have considered a case with multiple well-separated local minima in the potential $H_Z$, where one minimum (not the global minimum) is wider and so its energy drops more rapidly as a function of transverse field.
This is the setting which is worst for the adiabatic algorithm, but some nontrivial speedup is still found for the short path algorithm.
The speedup is modest, but this may be because we have taken such a large $\delta_w$.  For smaller $\delta_w$ (which also may be more realistic) the speedup becomes bigger.

For the case of a piecewise linear potential, a Grover speedup of a greedy classical algorithm works best.  However, we find that adding fluctuations to the potential (which will defeat this simple algorithm) has little effect on the short path.  Of course, this is not to be interpreted as implying that no classical algorithm can do well in this case.  Rather, it is a simple case with many minima that can still be studied numerically at large sizes.

The potential $H_Z$ that we consider is not one of those for which the previous proofs on the short path algorithm work since it cannot be written as homogeneous polynomial in variables $Z_i$.  This means that $H_Z$, averaged over points at given Hamming distance from the global minimum, may behave in a more complicated way than expected; the proofs regarding the short path algorithm rely heavily on properties of this average.
However, still some speedup is found.  Thus, this suggests applying the algorithm more broadly.

{\it Acknowledgments---} I thank D. Wecker for useful comments.

\bibliography{toy-ref}
\end{document}